%
% set these sizes for use with gstex under linux
%
% comment out the sizes for use with gsvga under linux
% also comment them out if running xdvi under X
%
\magnification = \magstep 1
% \hsize=16 true cm
% \vsize=10 true cm

\baselineskip=15pt plus 1pt minus 1pt
\font\secfnt=cmss10
%\font\vecfnt=cmbsy10

\font\titlefnt=cmssbx10 scaled \magstep 1
\def\title#1{\centerline{{\titlefnt #1}}\medskip}

\def\author#1{\smallskip\centerline{#1}\medskip}
\def\address#1{\centerline{#1}}
\def\date#1{\smallskip\centerline{{\it #1}}\smallskip}

\def\abstract#1{\par\vskip\normalbaselineskip\par
    {\baselineskip=\normalbaselineskip
    \parindent=0 pt
    {\hfill\vbox{\hsize= 11 cm  #1  }\hfill}}
    \bigskip}

\def\section#1{\bigskip\centerline{{\secfnt #1}}\medskip}

% auto-numbering -- equations

\newcount\eqncnt
\eqncnt=0
\def\eqprefix{}
\def\eqn{\global\advance\eqncnt by 1 {\rm(\eqprefix\the\eqncnt)}}
\def\eqname#1{\eqn\xdef#1{\eqprefix\the\eqncnt}}

% auto-numbering -- references

\newcount\refcnt
\refcnt=0
\def\ref#1.#2\par{\global\advance\refcnt by 1\xdef#1{\the\refcnt}}

% auto-numbering -- figures

\newcount\figcnt
\figcnt=0
\def\fig#1.#2\par{\global\advance\figcnt by 1\xdef#1{\the\figcnt}}

\def\etal{{\it et al}}
%\def\dh{{\it d.h.\ }}
%\def\zb{{\it z.B.\ }}
%\def\bzw{{\it bzw.\ }}

% inline fractions

% \def\frac#1/#2{{\scriptstyle{#1\over #2} }}
\def\half{ {1\over 2} }

\def\grapprox{\mathop{\lower.5ex \hbox{$\buildrel{\fivesy >}\over{\fivesy\sim}$}} \nolimits}
\def\lsapprox{\mathop{\lower.5ex \hbox{$\buildrel{\fivesy <}\over{\fivesy\sim}$}} \nolimits}
\def\grls{\mathop{\lower.5ex \hbox{$\buildrel{\fivesy >}\over{\fivesy <}$}} \nolimits}

\def\vec#1{{\bf #1}}

\def\avg#1{\left\langle #1 \right\rangle}
\def\abs#1{\left\vert #1 \right\vert}

\def\LBR{\left\lbrace}
\def\RBR{\right\rbrace}

\def\LP{\left (}
\def\RP{\right )}
\def\qq{\qquad\qquad}

\def\pt{\partial}

\def\pxx#1{{\partial #1\over\partial x}}

\def\pww#1{{\partial #1\over\partial w}}

\def\ptt#1{{\partial #1\over\partial t}}

\def\dtt#1{{d #1\over dt}}

\def\ppx#1{\partial #1/\partial x}
\def\ppy#1{\partial #1/\partial y}

\def\grad{\nabla}
\def\cross{{\bf \times}}
\def\div{\grad\cdot}

\def\dpl{\grad_\parallel}

\def\ddpp{\grad_\perp^2}

\def\bunit{\vec{b}}

\def\to{\rightarrow}

\def\cm{\,{\rm cm}}

\def\invcc{\,{\rm cm}^{-3}}

\def\eV{\,{\rm eV}}

\def\tesla{\,{\rm T}}

\def\Bdel{\vec B\cdot\grad}

\def\bdot{\vec B\cdot}

\def\vexb{\vec v_E}

\def\vedl{\vexb\cdot\grad}

\def\wpl{w_\parallel}

\def\Jpl{J_\parallel}

\def\Apl{A_\parallel}

\def\kpl{k_\parallel}

\def\kpp{k_\perp}

\def\kkpp{k_\perp^2}

\def\Dpl{D_\parallel}
\def\Dpp{\Delta_\perp}

\def\Rpl{R_\parallel}

\def\qepl{q_e{}_\parallel}
\def\qipl{q_i{}_\parallel}

\def\wpe{\omega_{pe}}
\def\wpi{\omega_{pi}}

\def\kappacv{{\cal K}}
\def\wcv{{\omega_B}}

\def\rs{\rho_s}

\def\Lpp{L_\perp}

\def\npl{\eta_\parallel}

\def\gaml{\Gamma_l}

\def\ptb{\widetilde}

\def\phifl{\widetilde\phi}

\def\nefl{\widetilde n_e}
\def\nifl{\widetilde n_i}

\def\pefl{\widetilde p_e}

\def\Bfl{\widetilde \vec B}

\def\ufl{\widetilde u_\parallel}
\def\vorfl{\grad_\perp^2\phifl}

\def\Afl{\ptb A_\parallel}
\def\Jfl{\ptb J_\parallel}

\def\teplfl{\widetilde T_e{}_\parallel}
\def\teppfl{\widetilde T_e{}_\perp}
\def\qeplfl{\widetilde q_e{}_\parallel}
\def\qeppfl{\widetilde q_e{}_\perp}
\def\tiplfl{\widetilde T_i{}_\parallel}
\def\tippfl{\widetilde T_i{}_\perp}
\def\qiplfl{\widetilde q_i{}_\parallel}
\def\qippfl{\widetilde q_i{}_\perp}

\def\tepl{ T_e{}_\parallel}
\def\tepp{ T_e{}_\perp}
\def\qepl{ q_e{}_\parallel}
\def\qepp{ q_e{}_\perp}
\def\tipl{ T_i{}_\parallel}
\def\tipp{ T_i{}_\perp}
\def\qipl{ q_i{}_\parallel}
\def\qipp{ q_i{}_\perp}

\def\peplfl{\widetilde p_e{}_\parallel}
\def\peppfl{\widetilde p_e{}_\perp}
\def\piplfl{\widetilde p_i{}_\parallel}
\def\pippfl{\widetilde p_i{}_\perp}

\def\pepl{ p_e{}_\parallel}
\def\pepp{ p_e{}_\perp}
\def\pipl{ p_i{}_\parallel}
\def\pipp{ p_i{}_\perp}

% normalised fluctuations

\def\shat{\hat s}
\def\bhat{\hat\beta}
\def\muhat{\hat\mu}
\def\epss{\hat\epsilon}

\def\prl#1{{\it Phys. Rev. Lett.} {\secfnt #1}}

\def\pf#1{{\it Phys. Fluids} {\secfnt #1}}

\def\pfb#1{{\it Phys. Fluids B} {\secfnt #1}}
\def\physp#1{{\it Phys. Plasmas} {\secfnt #1}}
\def\nf#1{{\it Nucl. Fusion} {\secfnt #1}}
\def\njp#1{{\it New J. Phys.} {\secfnt #1}}

\def\ppcf#1{{\it Plasma Phys. Contr. Fusion} {\secfnt #1}}

\def\revpp#1{{\it Rev. Plasma Phys.} {\secfnt #1}}
\def\iaea#1#2{in {\it Plasma Physics and Controlled Nuclear Fusion
    Research #1}, Proceedings of the #2th International Conference}

\def\aa#1{{\it Astron. Astrophys.} {\secfnt #1}}

\def\temp{1.34}%
\let\tempp=\relax
\expandafter\ifx\csname psboxversion\endcsname\relax
  \message{PSBOX(\temp) loading}%
\else
    \ifdim\temp cm>\psboxversion cm
      \message{PSBOX(\temp) loading}%
    \else
      \message{PSBOX(\psboxversion) is already loaded: I won't load
        PSBOX(\temp)!}%
      \let\temp=\psboxversion
      \let\tempp= 
    \fi
\fi
\tempp
\let\psboxversion=\temp
\catcode`\@=11
% Every macro likes a little privacy...
%
%Trying to tame the variety of \special commands for Postscript: the
%  universal internal command \PSspeci@l##1##2 takes ##1 to be the
%  filename and ##2 to be the integer scale factor*1000 (as for usual
%   TeX \scale commands)
%
\def\psfortextures{%     For TeXtures on the Macintosh
%-----------------
\def\PSspeci@l##1##2{%
\special{illustration ##1\space scaled ##2}%
}}%
\def\psfordvitops{%      For the DVItoPS converter on IBM mainframes
%----------------
\def\PSspeci@l##1##2{%
\special{dvitops: import ##1\space \the\drawingwd \the\drawinght}%
}}%
\def\psfordvips{%      For DVIPS converter on VAX, UNIX and PC's
%--------------
\def\PSspeci@l##1##2{%
%    \special{/@scaleunit 1000 def}% never read dox without trying!
\d@my=0.1bp \d@mx=\drawingwd \divide\d@mx by\d@my% BUG! for large \drawingwd
\includegraphics{##1\space}}}%
\def\psforoztex{%        For the OzTeX shareware on the Macintosh
%--------------
\def\PSspeci@l##1##2{%
\special{##1 \space
      ##2 1000 div dup scale
      \number-\psllx\space \number-\pslly\space translate
}}}%
\def\psfordvitps{%       From the UNIX TeXPS package, vers.>3.12
%---------------
% Convert a dimension into the number \psn@sp (in scaled points)
\def\psdimt@n@sp##1{\d@mx=##1\relax\edef\psn@sp{\number\d@mx}}
\def\PSspeci@l##1##2{%
% psfig.psr contains the def of "startTexFig": if you can locate it
% and include the correct pathname, it should work
\special{dvitps: Include0 "psfig.psr"}% contains def of "startTexFig"
\psdimt@n@sp{\drawingwd}
\special{dvitps: Literal "\psn@sp\space"}
\psdimt@n@sp{\drawinght}
\special{dvitps: Literal "\psn@sp\space"}
\psdimt@n@sp{\psllx bp}
\special{dvitps: Literal "\psn@sp\space"}
\psdimt@n@sp{\pslly bp}
\special{dvitps: Literal "\psn@sp\space"}
\psdimt@n@sp{\psurx bp}
\special{dvitps: Literal "\psn@sp\space"}
\psdimt@n@sp{\psury bp}
\special{dvitps: Literal "\psn@sp\space startTexFig\space"}
\special{dvitps: Include1 "##1"}
\special{dvitps: Literal "endTexFig\space"}
}}%
\def\psfordvialw{%   Try for dvialw, a UNIX public domain
%---------------
\def\PSspeci@l##1##2{
\special{language "PostScript",
position = "bottom left",
literal "  \psllx\space \pslly\space translate
  ##2 1000 div dup scale
  -\psllx\space -\pslly\space translate",
include "##1"}
}}%
\def\psforptips{%   For MS-DOS; LUOMA@brandeis.bitnet
%---------------
\def\PSspeci@l##1##2{{
\d@mx=\psurx bp
\advance \d@mx by -\psllx bp
\divide \d@mx by 1000\multiply\d@mx by \xscale
\incm{\d@mx}
\let\tmpx\dimincm
\d@my=\psury bp
\advance \d@my by -\pslly bp
\divide \d@my by 1000\multiply\d@my by \xscale
\incm{\d@my}
\let\tmpy\dimincm
\d@mx=-\psllx bp
\divide \d@mx by 1000\multiply\d@mx by \xscale
\d@my=-\pslly bp
\divide \d@my by 1000\multiply\d@my by \xscale
\at(\d@mx;\d@my){\special{ps:##1 x=\tmpx, y=\tmpy}}
}}}%
\def\psonlyboxes{%     Draft-like behaviour if none of the others works
%---------------
\def\PSspeci@l##1##2{%
\at(0cm;0cm){\boxit{\vbox to\drawinght
  {\vss\hbox to\drawingwd{\at(0cm;0cm){\hbox{({\tt##1})}}\hss}}}}
}}%
\def\psloc@lerr#1{%
\let\savedPSspeci@l=\PSspeci@l%
\def\PSspeci@l##1##2{%
\at(0cm;0cm){\boxit{\vbox to\drawinght
  {\vss\hbox to\drawingwd{\at(0cm;0cm){\hbox{({\tt##1}) #1}}\hss}}}}
\let\PSspeci@l=\savedPSspeci@l% restore normal output for other figs!
}}%
%\def\psfor...  add your own!
%
% Some common defs
%
\newread\pst@mpin
\newdimen\drawinght\newdimen\drawingwd
\newdimen\psxoffset\newdimen\psyoffset
\newbox\drawingBox
\newcount\xscale \newcount\yscale \newdimen\pscm\pscm=1cm
\newdimen\d@mx \newdimen\d@my
\newdimen\pswdincr \newdimen\pshtincr
\let\ps@nnotation=\relax
{\catcode`\|=0 |catcode`|\=12 |catcode`|%=12 |catcode`~=12
|catcode`#=12 |catcode`*=14
|xdef|backslashother{\}*
|xdef|percentother{%}*
|xdef|tildeother{~}*
|xdef|sharpother{#}*
}%
% useful to display special chars in \tt; fails for \,#,%
\def\R@moveMeaningHeader#1:->{}%
\def\uncatcode#1{%
\edef#1{\expandafter\R@moveMeaningHeader\meaning#1}}%
\def\execute#1{#1}% NOT stupid: cs in #1 are then identified BEFORE execution
\def\psm@keother#1{\catcode`#112\relax}% borrowed from latex
\def\executeinspecs#1{%
\execute{\begingroup\let\do\psm@keother\dospecials\catcode`\^^M=9#1\endgroup}}%
\def\@mpty{}%
% \if\matchin#1#2<=> \iftrue if #1 contains #2, <=>\iffalse otherwise:
% \if\matchexpin: idem, but #1 & #2 are first fully expanded (no \if
% inside!)
% \tmpa & \tmpb contain what's before and after the occurence of #2
\def\matchexpin#1#2{
  \fi%
%\message{(#1>#2)}
  \edef\tmpb{{#2}}%
  \expandafter\makem@tchtmp\tmpb%
  \edef\tmpa{#1}\edef\tmpb{#2}%
  \expandafter\expandafter\expandafter\m@tchtmp\expandafter\tmpa\tmpb\endm@tch%
  \if\match%
}%
\def\matchin#1#2{%
  \fi%
  \makem@tchtmp{#2}%
  \m@tchtmp#1#2\endm@tch%
  \if\match%
}%
\def\makem@tchtmp#1{\def\m@tchtmp##1#1##2\endm@tch{%
  \def\tmpa{##1}\def\tmpb{##2}\let\m@tchtmp=\relax%
  \ifx\tmpb\@mpty\def\match{YN}%
  \else\def\match{YY}\fi%
}}%
% converts any dimen in cm, with 1E-4 cm precision
\def\incm#1{{\psxoffset=1cm\d@my=#1
 \d@mx=\d@my
  \divide\d@mx by \psxoffset
  \xdef\dimincm{\number\d@mx.}
  \advance\d@my by -\number\d@mx cm
  \multiply\d@my by 100
 \d@mx=\d@my
  \divide\d@mx by \psxoffset
  \edef\dimincm{\dimincm\number\d@mx}
  \advance\d@my by -\number\d@mx cm
  \multiply\d@my by 100
 \d@mx=\d@my
  \divide\d@mx by \psxoffset
  \xdef\dimincm{\dimincm\number\d@mx}
}}%
%
%  \ReadPSize{PSfilename} reads the dimensions of a PostScript drawing
%      and stores it in \drawinght(wd)
\newif\ifNotB@undingBox
\newhelp\PShelp{Proceed: you'll have a 5cm square blank box instead of
your graphics (Jean Orloff).}%
\def\s@tsize#1 #2 #3 #4\@ndsize{
  \def\psllx{#1}\def\pslly{#2}%
  \def\psurx{#3}\def\psury{#4}%  needed by a crazyness of dvips!
  \ifx\psurx\@mpty\NotB@undingBoxtrue% this is not a valid one!
  \else
    \drawinght=#4bp\advance\drawinght by-#2bp
    \drawingwd=#3bp\advance\drawingwd by-#1bp
%  !Units related by crazy factors as bp/pt=72.27/72 should be BANNED!
  \fi
  }%
\def\sc@nBBline#1:#2\@ndBBline{\edef\p@rameter{#1}\edef\v@lue{#2}}%
\def\g@bblefirstblank#1#2:{\ifx#1 \else#1\fi#2}%
{\catcode`\%=12
\xdef\B@undingBox{%%BoundingBox}}%
%% is not a true comment in PostScript, even if % is!
\def\ReadPSize#1{
 \readfilename#1\relax
 \let\PSfilename=\lastreadfilename
 \openin\pst@mpin=#1\relax
 \ifeof\pst@mpin \errhelp=\PShelp
   \errmessage{I haven't found your postscript file (\PSfilename)}%
   \psloc@lerr{was not found}%
   \s@tsize 0 0 142 142\@ndsize
   \closein\pst@mpin
 \else
% each entry in \GlobalInputList should be unique
   \if\matchexpin{\GlobalInputList}{, \lastreadfilename}%
   \else\xdef\GlobalInputList{\GlobalInputList, \lastreadfilename}%
     \immediate\write\psbj@inaux{\lastreadfilename,}%
   \fi%
   \loop
     \executeinspecs{\catcode`\ =10\global\read\pst@mpin to\n@xtline}%
     \ifeof\pst@mpin
       \errhelp=\PShelp
       \errmessage{(\PSfilename) is not an Encapsulated PostScript File:
           I could not find any \B@undingBox: line.}%
       \edef\v@lue{0 0 142 142:}%
       \psloc@lerr{is not an EPSFile}%
       \NotB@undingBoxfalse
     \else
       \expandafter\sc@nBBline\n@xtline:\@ndBBline
       \ifx\p@rameter\B@undingBox\NotB@undingBoxfalse
         \edef\t@mp{%
           \expandafter\g@bblefirstblank\v@lue\space\space\space}%
         \expandafter\s@tsize\t@mp\@ndsize
       \else\NotB@undingBoxtrue
       \fi
     \fi
   \ifNotB@undingBox\repeat
   \closein\pst@mpin
 \fi
\message{#1}%
}%
%
% \psboxto(xdim;ydim){psfilename}: you specify the dimensions and
%    TeX uniformly scales to fit the largest one. If xdim=0pt, the
%    scale is fully determined by ydim and vice versa.
%    Notice: psboxes are a real vboxes; couldn't take hbox otherwise all
%    indentation and all cr's would be interpreted as spaces (hugh!).
%
\def\psboxto(#1;#2)#3{\vbox{
   \ReadPSize{#3}%
   \divide\drawingwd by 1000
   \divide\drawinght by 1000
   \d@mx=#1
   \ifdim\d@mx=0pt\xscale=1000
         \else \xscale=\d@mx \divide \xscale by \drawingwd\fi
   \d@my=#2
   \ifdim\d@my=0pt\yscale=1000
         \else \yscale=\d@my \divide \yscale by \drawinght\fi
   \ifnum\yscale=1000
         \else\ifnum\xscale=1000\xscale=\yscale
                    \else\ifnum\yscale<\xscale\xscale=\yscale\fi
              \fi
   \fi
   \divide\pswdincr by 1000 \multiply\pswdincr by \xscale
   \divide\pshtincr by 1000 \multiply\pshtincr by \xscale
   \divide\psxoffset by1000 \multiply\psxoffset by\xscale
   \divide\psyoffset by1000 \multiply\psyoffset by\xscale
   \global\divide\pscm by 1000
   \global\multiply\pscm by\xscale
   \multiply\drawingwd by\xscale \multiply\drawinght by\xscale
   \ifdim\d@mx=0pt\d@mx=\drawingwd\fi
   \ifdim\d@my=0pt\d@my=\drawinght\fi
   \message{scaled \the\xscale}%
 \hbox to\d@mx{\hss\vbox to\d@my{\vss
   \global\setbox\drawingBox=\hbox to 0pt{\kern\psxoffset\vbox to 0pt{
      \kern-\psyoffset
      \PSspeci@l{\PSfilename}{\the\xscale}%
      \vss}\hss\ps@nnotation}%
   \advance\pswdincr by \drawingwd
   \advance\pshtincr by \drawinght
   \global\wd\drawingBox=\the\pswdincr
   \global\ht\drawingBox=\the\pshtincr
   \baselineskip=0pt
   \copy\drawingBox
 \vss}\hss}%
  \global\psxoffset=0pt
  \global\psyoffset=0pt
  \global\pswdincr=0pt
  \global\pshtincr=0pt % These are local to one figure
  \global\pscm=1cm %should not be necessary
  \global\drawingwd=\drawingwd
  \global\drawinght=\drawinght
}}%
%
% \psboxscaled{scalefactor*1000}{PSfilename} allows to bypass the
%   rounding errors of TeX integer divisions for situations where the
%   TeX box should fit the original BoundingBox with a precision better
%   than 1/1000.
%
\def\psboxscaled#1#2{\vbox{
  \ReadPSize{#2}%
  \xscale=#1
  \message{scaled \the\xscale}%
  \advance\drawingwd by\pswdincr\advance\drawinght by\pshtincr
  \divide\pswdincr by 1000 \multiply\pswdincr by \xscale
  \divide\pshtincr by 1000 \multiply\pshtincr by \xscale
  \divide\psxoffset by1000 \multiply\psxoffset by\xscale
  \divide\psyoffset by1000 \multiply\psyoffset by\xscale
  \divide\drawingwd by1000 \multiply\drawingwd by\xscale
  \divide\drawinght by1000 \multiply\drawinght by\xscale
  \global\divide\pscm by 1000
  \global\multiply\pscm by\xscale
  \global\setbox\drawingBox=\hbox to 0pt{\kern\psxoffset\vbox to 0pt{
     \kern-\psyoffset
     \PSspeci@l{\PSfilename}{\the\xscale}%
     \vss}\hss\ps@nnotation}%
  \advance\pswdincr by \drawingwd
  \advance\pshtincr by \drawinght
  \global\wd\drawingBox=\the\pswdincr
  \global\ht\drawingBox=\the\pshtincr
  \baselineskip=0pt
  \copy\drawingBox
  \global\psxoffset=0pt
  \global\psyoffset=0pt
  \global\pswdincr=0pt
  \global\pshtincr=0pt % These are local to one figure
  \global\pscm=1cm
  \global\drawingwd=\drawingwd
  \global\drawinght=\drawinght
}}%
%
%  \psbox{PSfilename} makes a TeX box having the minimal size to
%      enclose the picture
\def\psbox#1{\psboxscaled{1000}{#1}}%
%------------------------------------------------------
%  \joinfiles file1, file2, ...n \into joinedfilename .
%     makes one file out of many
%  \splitfile joinedfilename
%     the opposite
\newif\ifn@teof\n@teoftrue
\newif\ifc@ntrolline
\newif\ifmatch
\newread\j@insplitin
\newwrite\j@insplitout
\newwrite\psbj@inaux
\immediate\openout\psbj@inaux=psbjoin.aux
\immediate\write\psbj@inaux{\string\joinfiles}%
\immediate\write\psbj@inaux{\jobname,}%
%
% INPUT REDEFINITION
%
% works if #1 is a single character
\def\toother#1{\ifcat\relax#1\else\expandafter%
  \toother@ux\meaning#1\endtoother@ux\fi}%
\def\toother@ux#1 #2#3\endtoother@ux{\def\tmp{#3}%
  \ifx\tmp\@mpty\def\tmp{#2}\let\next=\relax%
  \else\def\next{\toother@ux#2#3\endtoother@ux}\fi%
\next}%
%
% \readfilename defs:
%
\let\readfilenamehook=\relax
\def\re@d{\expandafter\re@daux}% spares typing 10 \expandafter's...
\def\re@daux{\futurelet\nextchar\stopre@dtest}%
\def\re@dnext{\xdef\lastreadfilename{\lastreadfilename\nextchar}%
  \afterassignment\re@d\let\nextchar}%
\def\stopre@d{\egroup\readfilenamehook}%
\def\stopre@dtest{%
  \ifcat\nextchar\relax\let\nextread\stopre@d
  \else
    \ifcat\nextchar\space\def\nextread{%
      \afterassignment\stopre@d\chardef\nextchar=`}%
    \else\let\nextread=\re@dnext
      \toother\nextchar
      \edef\nextchar{\tmp}%
    \fi
  \fi\nextread}%
\def\readfilename{\vbox\bgroup%
  \let\\=\backslashother \let\%=\percentother \let\~=\tildeother
  \let\#=\sharpother \xdef\lastreadfilename{}%
  \re@d}%
%
% redefines \input using \readfilename
%
\xdef\GlobalInputList{\jobname}%
\def\psnewinput{%
  \def\readfilenamehook{% each entry in \GlobalInputList should be unique
    \if\matchexpin{\GlobalInputList}{, \lastreadfilename}%
    \else\xdef\GlobalInputList{\GlobalInputList, \lastreadfilename}%
      \immediate\write\psbj@inaux{\lastreadfilename,}%
    \fi%
    \ps@ldinput\lastreadfilename\relax%
    \let\readfilenamehook=\relax%
  }\readfilename%
}%
\expandafter\ifx\csname @@input\endcsname\relax    % then Plain
  \immediate\let\ps@ldinput=\input\def\input{\psnewinput}%
\else
  \immediate\let\ps@ldinput=\@@input
  \def\@@input{\psnewinput}%
\fi%
\def\nowarnopenout{%
 \def\warnopenout##1##2{%
   \readfilename##2\relax
   \message{\lastreadfilename}%
   \immediate\openout##1=\lastreadfilename\relax}}%
\def\warnopenout#1#2{%
 \readfilename#2\relax
 \def\t@mp{TrashMe,psbjoin.aux,psbjoint.tex,}\uncatcode\t@mp
 \if\matchexpin{\t@mp}{\lastreadfilename,}%
 \else
   \immediate\openin\pst@mpin=\lastreadfilename\relax
   \ifeof\pst@mpin
     \else
     \errhelp{If the content of this file is so precious to you, abort (ie
press x or e) and rename it before retrying.}%
     \errmessage{I'm just about to replace your file named \lastreadfilename}%
   \fi
   \immediate\closein\pst@mpin
 \fi
 \message{\lastreadfilename}%
 \immediate\openout#1=\lastreadfilename\relax}%
% % will have an unusual catcode below; use * instead
%\vbox
{\catcode`\%=12\catcode`\*=14
\gdef\splitfile#1{*
 \readfilename#1\relax
 \immediate\openin\j@insplitin=\lastreadfilename\relax
 \ifeof\j@insplitin
   \message{! I couldn't find and split \lastreadfilename!}*
 \else
   \immediate\openout\j@insplitout=TrashMe
   \message{< Splitting \lastreadfilename\space into}*
   \loop
     \ifeof\j@insplitin
       \immediate\closein\j@insplitin\n@teoffalse
     \else
       \n@teoftrue
       \executeinspecs{\global\read\j@insplitin to\spl@tinline\expandafter
         \ch@ckbeginnewfile\spl@tinline%Beginning-Of-File-Named:%\endcheck}*
       \ifc@ntrolline
       \else
         \toks0=\expandafter{\spl@tinline}*
         \immediate\write\j@insplitout{\the\toks0}*
       \fi
     \fi
   \ifn@teof\repeat
   \immediate\closeout\j@insplitout
 \fi\message{>}*
}*
\gdef\ch@ckbeginnewfile#1%Beginning-Of-File-Named:#2%#3\endcheck{*
 \def\t@mp{#1}*
 \ifx\@mpty\t@mp
   \def\t@mp{#3}*
   \ifx\@mpty\t@mp
     \global\c@ntrollinefalse
   \else
     \immediate\closeout\j@insplitout
     \warnopenout\j@insplitout{#2}*
     \global\c@ntrollinetrue
   \fi
 \else
   \global\c@ntrollinefalse
 \fi}*
\gdef\joinfiles#1\into#2{*
 \message{< Joining following files into}*
 \warnopenout\j@insplitout{#2}*
 \message{:}*
 {*
 \edef\w@##1{\immediate\write\j@insplitout{##1}}*
\w@{% This collection of files was produced with CERN psbox package}*
\w@{% To decompose and tex it:}*
\w@{%-save this with a filename CONTAINING ONLY LETTERS and a .TEX}*
\w@{% extension (say, JOINTFIL.TEX), in some uncrowded directory;}*
\w@{%-make sure you can \string\input\space psbox.tex (version>=1.3);}*
\w@{%  (else ftp cs.nyu.edu(=128.122.140.24):pub/TeX/psbox/, then get}*
\w@{%  and tex the file psboxall.tex; more info in psbREAD.ME)}*
\w@{%-tex JOINTFIL.TEX using Plain, or LaTeX, or whatever is needed by}*
\w@{%  the first file in the joining (after splitting JOINTFIL.TEX into}*
\w@{%  it's constituents, TeX will try to process it as it stands).}*
\w@{\string\input\space psbox.tex}*
\w@{\string\splitfile{\string\jobname}}*
\w@{\string\let\string\autojoin=\string\relax}*
}*
 \expandafter\tre@tfilelist#1, \endtre@t
 \immediate\closeout\j@insplitout
 \message{>}*
}*
\gdef\tre@tfilelist#1, #2\endtre@t{*
 \readfilename#1\relax
 \ifx\@mpty\lastreadfilename
 \else
   \immediate\openin\j@insplitin=\lastreadfilename\relax
   \ifeof\j@insplitin
     \errmessage{I couldn't find file \lastreadfilename}*
   \else
     \message{\lastreadfilename}*
     \immediate\write\j@insplitout{%Beginning-Of-File-Named:\lastreadfilename}*
     \executeinspecs{\global\read\j@insplitin to\oldj@ininline}*
     \loop
       \ifeof\j@insplitin\immediate\closein\j@insplitin\n@teoffalse
       \else\n@teoftrue
         \executeinspecs{\global\read\j@insplitin to\j@ininline}*
         \toks0=\expandafter{\oldj@ininline}*
         \let\oldj@ininline=\j@ininline
         \immediate\write\j@insplitout{\the\toks0}*
       \fi
     \ifn@teof
     \repeat
   \immediate\closein\j@insplitin
   \fi
   \tre@tfilelist#2, \endtre@t
 \fi}*
}%
% To be put at the end of a file, for making a tar-like file containing
%   everything it used.
\def\autojoin{%
 \immediate\write\psbj@inaux{\string\into{psbjoint.tex}}%
 \immediate\closeout\psbj@inaux
 \expandafter\joinfiles\GlobalInputList\into{psbjoint.tex}%
}%
%----------------------------------------------------------------
%  Annotations & Captions etc...
%
%
% \centinsert{anybox} is just a centered \midinsert, but is included as
%    people barely use the original inserts from TeX.
%
\def\centinsert#1{\midinsert\line{\hss#1\hss}\endinsert}%
\def\psannotate#1#2{\vbox{%
  \def\ps@nnotation{#2\global\let\ps@nnotation=\relax}#1}}%
\def\pscaption#1#2{\vbox{%
   \setbox\drawingBox=#1
   \copy\drawingBox
   \vskip\baselineskip
   \vbox{\hsize=\wd\drawingBox\setbox0=\hbox{#2}%
     \ifdim\wd0>\hsize
       \noindent\unhbox0\tolerance=5000
    \else\centerline{\box0}%
    \fi
}}}%
% for compatibility with older versions, but \psfig is a bad name!
%\def\psfig#1#2#3{\pscaption{\psannotate{#1}{#2}}{#3}}
%\def\psfigurebox#1#2#3{\pscaption{\psannotate{\psbox{#1}}{#2}}{#3}}
%
% \at(#1;#2)#3 puts #3 at #1-higher and #2-right of the current
%    position without moving it (to be used in annotations).
\def\at(#1;#2)#3{\setbox0=\hbox{#3}\ht0=0pt\dp0=0pt
  \rlap{\kern#1\vbox to0pt{\kern-#2\box0\vss}}}%
%
% \gridfill(ht;wd) makes a 1cm*1cm grid of ht by wd whose lower-left
%   corner is the current point
\newdimen\gridht \newdimen\gridwd
\def\gridfill(#1;#2){%
  \setbox0=\hbox to 1\pscm
  {\vrule height1\pscm width.4pt\leaders\hrule\hfill}%
  \gridht=#1
  \divide\gridht by \ht0
  \multiply\gridht by \ht0
  \gridwd=#2
  \divide\gridwd by \wd0
  \multiply\gridwd by \wd0
  \advance \gridwd by \wd0
  \vbox to \gridht{\leaders\hbox to\gridwd{\leaders\box0\hfill}\vfill}}%
%
% Useful to measure where to put annotations
\def\fillinggrid{\at(0cm;0cm){\vbox{%
  \gridfill(\drawinght;\drawingwd)}}}%
%
% \textleftof\anybox: Sample text\endtext
%   inserts "Sample text" on the left of \anybox ie \vbox, \psbox.
%   \textrightof is the symmetric (not documented, too uggly)
% Welcome any suggestion about clean wraparound macros from
%   TeXhackers reading this
%
\def\textleftof#1:{%
  \setbox1=#1
  \setbox0=\vbox\bgroup
    \advance\hsize by -\wd1 \advance\hsize by -2em}%
\def\textrightof#1:{%
  \setbox0=#1
  \setbox1=\vbox\bgroup
    \advance\hsize by -\wd0 \advance\hsize by -2em}%
\def\endtext{%
  \egroup
  \hbox to \hsize{\valign{\vfil##\vfil\cr%
\box0\cr%
\noalign{\hss}\box1\cr}}}%
%
% \frameit{\thick}{\skip}{\anybox}
%    draws with thickness \thick a box around \anybox, leaving \skip of
%    blank around it. eg \frameit{0.5pt}{1pt}{\hbox{hello}}
% \boxit{\anybox} is a shortcut.
\def\frameit#1#2#3{\hbox{\vrule width#1\vbox{%
  \hrule height#1\vskip#2\hbox{\hskip#2\vbox{#3}\hskip#2}%
        \vskip#2\hrule height#1}\vrule width#1}}%
\def\boxit#1{\frameit{0.4pt}{0pt}{#1}}%
\catcode`\@=12 % cs containing @ are unreachable
%
% CUSTOMIZE YOUR DEFAULT DRIVER:
%    Uncomment the line corresponding to your TeX system:
%\psfortextures%     For TeXtures on the Macintosh
%\psforoztex   %     For OzTeX shareware on the Macintosh
%\psfordvitops %     For the DVItoPS converter for TeX on IBM mainframes
 \psfordvips   %     For DVIPS converter on VAX and UNIX
%\psfordvitps  %     For dvitps from TeXPS package under UNIX
%\psfordvialw  %     For dvialw, UNIX public domain
%\psonlyboxes  %     Blank Boxes (when all else fails).

\def\Dpp{\Delta_\perp}
\def\Dpl{\Delta_\parallel}
\def\Kpl{K_\parallel}

\def\vv{\vec v}
\def\uu{\vec u}
\def\xx{\vec x}
\def\ww{\vec w}
\def\ee{\vec E}
\def\bb{\vec B}
\def\ff{\vec F}
\def\jj{\vec J}
\def\qq{\vec q}
\def\aa{\vec A}
\def\kk{\vec k}

\def\qeperp{\vec q_e{}_\perp}
\def\qiperp{\vec q_i{}_\perp}
\def\Pipl{\Pi_\parallel}

\def\bdot{\bunit\cdot}

\def\wfl{\ptb w}
\def\Wfl{\ptb W}
\def\Bfl{\ptb B}
\def\Rfl{\ptb \Rpl}
\def\bbfl{\ptb\bb}
\def\bbpp{\ptb\bb_\perp}
\def\uufl{\ptb\uu}
\def\jjfl{\ptb\jj}
\def\rhofl{\ptb\rho}
\def\uxfl{\ptb u^x}
\def\uyfl{\ptb u^y}
\def\vexfl{\ptb v_E^x}
\def\veyfl{\ptb v_E^y}
\def\bflx{\ptb B^x}
\def\bfly{\ptb B^y}

\def\vprof{V}
\def\vprofp{V'}
\def\vprofpp{V''}

\def\wpe{\omega_{pe}}
\def\wpi{\omega_{pi}}

\def\drift{{c\over B^2}\vec B\cross}
\def\wL{\omega_L}

\def\lcorx{\lambda_x}
\def\lcory{\lambda_y}
\def\tcor{\tau_c}

\def\dalpha{\Delta}
\def\pxxmu#1{{\pt #1\over\pt x^\mu}}
\def\pyyk#1{{\pt #1\over\pt y_k}}
\def\ppyyk#1{{\pt^2 #1\over\pt y_k^2}}
\def\pww#1{{\pt #1\over\pt \wpl}}

\def\gamm{\Gamma_m}
\def\gaml{\Gamma_l}

\def\nefl{\widetilde n_e}
\def\teplfl{\widetilde T_e{}_\parallel}
\def\teppfl{\widetilde T_e{}_\perp}
\def\qeplfl{\widetilde q_e{}_\parallel}
\def\qeppfl{\widetilde q_e{}_\perp}
\def\nifl{\widetilde n_i}
\def\tiplfl{\widetilde T_i{}_\parallel}
\def\tippfl{\widetilde T_i{}_\perp}
\def\qiplfl{\widetilde q_i{}_\parallel}
\def\qippfl{\widetilde q_i{}_\perp}

\def\tepl{ T_e{}_\parallel}
\def\tepp{ T_e{}_\perp}
\def\qepl{ q_e{}_\parallel}
\def\qepp{ q_e{}_\perp}
\def\tipl{ T_i{}_\parallel}
\def\tipp{ T_i{}_\perp}
\def\qipl{ q_i{}_\parallel}
\def\qipp{ q_i{}_\perp}

\def\peplfl{\widetilde p_e{}_\parallel}
\def\peppfl{\widetilde p_e{}_\perp}
\def\piplfl{\widetilde p_i{}_\parallel}
\def\pippfl{\widetilde p_i{}_\perp}

\def\pepl{ p_e{}_\parallel}
\def\pepp{ p_e{}_\perp}
\def\pipl{ p_i{}_\parallel}
\def\pipp{ p_i{}_\perp}

\def\Eplfl{\ptb E_\parallel}

\def\chiv{\hat\chi}

\def\kkpp{k_\perp^2}

\def\uexb{\vec u_E}
\def\wexb{\vec w_E}

\def\uedl{\uexb\cdot\grad}
\def\wedl{\wexb\cdot\grad}

\def\section#1{
%	\par\vfill\eject
 	\centerline{{\secfnt #1}}}
\def\abstract#1{\vskip 2 true cm
 	\noindent{\hfill\vbox{\hsize = 14 true cm #1}}\hfill}

%   with un normalised eqs and more on energetics

\def\sc{\secfnt}

\title{Zonal Flows and Electromagnetic Drift Wave Turbulence}
\author{Bruce D. Scott}
\address{Max-Planck-IPP, EURATOM Association, 85748 Garching, Germany}
\date{Jul 2002}

\abstract{Detailed computations of tokamak edge turbulence in three
dimensional, globally consistent flux tube geometry show an inhibition
of the standard scenario in which zonal ExB flows generated by the
turbulence should lead to transport barrier formation.  It is found
by comparison to slab geometry and by analysis of the energetics
that the zonal flow energy is
depleted by toroidal coupling to the pressure through the geodesic
curvature.  Edge transport barriers would then depend on the physics of
the neoclassical equilibrium.}

\vfill

\leftline{PACS numbers: 52.25.Fi  91.25.Cw  52.30.-q  52.40.Nk}

\par\eject

{\sc Drift Wave Turbulence and Zonal Flows.}  Drift wave turbulence is
nonlinear, nonperiodic motion involving disturbances on a background
thermal gradient of a magnetised plasma and eddies of fluid like motion
in which the advecting velocity of all charged species is the ExB
velocity [\hasmim,\wakhas].  The disturbances in the electric field
implied by the presence of these eddies are caused by the tendency of
the electron dynamics to establish a force balance along the magnetic
field.  Pressure disturbances have their parallel gradients balanced by
a parallel electric field, whose static part is given by the parallel
gradient of the electrostatic potential.  This potential in turn is the
stream function for the ExB velocity in drift planes, which are locally
perpendicular to the magnetic field.  The turbulence is driven by the
background gradient, and the electron pressure and electrostatic
potential are coupled together through parallel currents.  Departures
from the static force balance are mediated primarily through
electromagnetic induction and resistive friction, but also the electron
inertia, which is not negligible [\dalfloc].  Further details are
provided by the temperatures, whose dynamics is very robust due to
nonlinear, time dependent Landau damping.  In a three dimensional,
toroidal flux surface geometry, the turbulence is characterised by a
nonlinear instability whose inherent vorticity is strong enough to
``supersede'' linear interchange instabilities, giving tokamak edge
turbulence a drift wave basic character [\focusdw].

Although this turbulence effects an unsteady transport through
nonlinear advection of the thermodynamic state variables, actual
modification of the profiles proceeds on the much slower transport time
scale, typically at least three orders of magnitude slower than the
turbulence even in steep gradient regions.  Such quasilinear
modification of the background (a three wave interaction involving a
wave and its complex conjugate driving changes in the background) also
occurs in the other variables, specifically the ExB vorticity, for which
the time scale is short enough for a self-consistent interaction with
the turbulence to affect the dynamics of the turbulence itself.  These
are the ``zonal flows'' [\zfterry], 
which are simply the ExB flows resulting from
disturbances in the electrostatic potential which are constant on a
given magnetic flux surface.  One can think of a zonal flow as a rigid
poloidal rotation of the entire flux surface.  These flows are important
because when and where they are sheared they can cause a local
suppression of the turbulence.  Suppression of turbulence by sheared
flows began by considering an imposed flow which is part of the
background [\biglari].
It was then pointed out that the process by which this suppression
occurs conserves energy, mainly involving a transfer of energy in three
wave interactions from
smaller scale eddies and the larger scale background
[\diamondkim], a variant of the more general inverse energy
cascade from smaller to larger scales in two dimensional, incompressible
turbulence [\kraichnan].
Suppression by
imposed ExB shear was then shown in computations to proceed
energetically [\sfdw].  Self
consistent interactions concern time dependent zonal flows which have
time and space scales comparable to or only somewhat larger than those
of the turbulence.  Their study as such [\hahm], followed global
scale computational studies showing them to be very important in
limiting the radial scale of the turbulence and consequently the
resulting transport [\zlin].  Their importance in fusion research lies
in the fact that zonal ExB flow (electric field) shear is thought
to underly the transition and maintenance of the H-mode
operation of tokamak confinement [\wagner].

We here examine the physics of the zonal flow/turbulence interaction in
a model of the turbulence which includes two important generalisations:
departures from the ``adiabatic'' state of perfect electron force
balance, and an electromagnetic character in that response, which allows
significant delays in the adiabatic response at larger perpendicular
scales, since the collisional response is through the parallel current,
while the inductive response is through the time dependence of the
parallel magnetic potential (by Ampere's law, the current is given by
the perpendicular Laplacian of that potential).  We will find that in
toroidal geometry, although the drive of the zonal flows via Reynolds
stress remains, the geodesic curvature of the magnetic field lines
couples the zonal flows to pressure sidebands with finite parallel
gradient.  These sidebands serve as a localised part of the general
source for the turbulence, as the free energy transfer in the pressure
disturbances is preferentially towards smaller scales.
The build up of strong, long lived ExB ``mean flow'' shear layers is
thereby inhibited, preventing the turbulence from self consistently
generating enough ExB shear to strongly reduce its own amplitude.  This
prevents the scenario in which the zonal flow drive process should lead
to transport barrier formation.  In slab geometry, the geodesic
curvature effect is absent and mean flows do develop, but we find by
inserting specifically this geodesic curvature effect that the toroidal
result is recovered (incidentally demonstrating the weakness of the
ballooning/interchange effect in the turbulence).  It is important to
note in this context that models of transport barrier formation by
Reynolds stress-induced self-generated flows rely on two-dimensional
slab geometry [\diamond], and they work well in such computations
[\varenna], but the three-dimensional toroidal result is rather
different as documented herein.

{\sc The DALF3 Model.}
The simplest three dimensional model of drift wave dynamics which
takes the self consistent adiabatic response into account and
allows it an electromagnetic character is a four field model in
toroidal flux tube geometry called DALF3 (the drift Alfv\'en model
[\dalfloc] but omitting the temperature dynamics).
The state variables are the electrostatic potential
$\phifl$ and the electron pressure $\pefl$, and the flux variables are
the parallel current $\Jfl$ and the parallel ion velocity $\ufl$, all
expressed as disturbances on the equilibrium which is a set of constant
parameters except where the ExB and magnetic nonlinearities operate on
the background gradients.  In the Ohm's law, electromagnetic induction, 
electron inertia, and resistive friction are all retained.
The adiabatic response is the reaction of the parallel current,
controlled by those three effects,
to the pressure/potential static force imbalance, acting to couple
$\pefl$ and $\phifl$ through the shear Alfv\'en dynamics.
The model equations in simplified sheared flux tube geometry are
$${n_eM_i c^2\over B^2}\dtt{}\ddpp\phifl 
	= B\dpl{\Jfl\over B} + \div\drift\grad\pefl
	\eqno\eqname\eqvorfl$$
$${1\over c}\ptt{\Afl}+{m_e\over n_e e^2}\dtt{\Jfl} + \npl\Jfl
	= {1\over n_e e}\dpl\LP p_e+\pefl\RP-\dpl\phifl
	\eqno\eqname\eqafl$$
$$\dtt{}\LP\pefl+p_e\RP
	= {T_e\over e}B\dpl{\Jfl\over B} 
	- p_e B\dpl{\ufl\over B} 
	- {1\over e}\div\drift\grad\pefl
	+ p_e\div\drift\grad\phifl
	\eqno\eqname\eqpefl$$
$$n_e M_i\dtt{\ufl} = -\dpl\LP p_e+\pefl\RP
	\eqno\eqname\equfl$$
with Ampere's law $\Jfl = -(c/4\pi)\ddpp\Afl$.
The ExB advective and parallel derivatives are given by
$$\dtt{}=\ptt{}+\drift\grad\phi\cdot\grad \qquad\qquad
	\dpl={1\over B}\Bdel - \drift\grad{1\over c}\Afl\cdot\grad
	\eqno\eqn$$
where $\vec B$ and $B$ are the equilibrium magnetic field and its
magnitude, and the combinations involving $\phifl$ and $\Afl$ give the
ExB and disturbed parallel derivatives, respectively, i.e., the
nonlinearities.
The flux tube geometry used is detailed elsewhere [\shifted], as is the
importance of global consistency which controls field line connection
[\fluxtube].
The standard normalisation is in terms of the drift scale
$\rs=c_s/\Omega_i$ and frequency $c_s/\Lpp$, where
$c_s^2=T_e/M_i$ and $\Omega_i=eB/M_i c$, and $\Lpp$ is the background
scale length for $p_e$.  
The parameters controlling the adiabatic response are
$\bhat=(c_s/\Lpp)^2(qR/v_A)^2$, and
$\muhat=(c_s/\Lpp)^2(qR/V_e)^2$, and $C=0.51(\nu_e\Lpp/c_s)\muhat$, 
reflecting the competition between perpendicular ExB turbulence and the
parallel dynamics, where the field line connection length is $2\pi qR$,
$V_e$ is the electron thermal velocity ($V_e^2=T_e/m_e$)
and the $0.51$ comes from the parallel resistivity,
$\npl=0.51 m_e\nu_e/n_e e^2$ [\brag].  The sound
waves are controlled by $\epss=(c_s/\Lpp)^2(qR/c_s)^2$, just the
parallel/perp scale ratio.  The effects of magnetic curvature (the
radius of curvature is $R$, the toroidal major radius), entering
through $\kappacv\equiv\div(c/B^2)\vec B\cross\grad$ are controlled by
$\wcv=2\Lpp/R$, which can be set independently --- slab geometry is
$\wcv=0$.
The coordinates are $\{x,y,s\}$, representing 
the down-gradient,
electron drift, and parallel directions, respectively.
The computations are set up exactly as detailed in [\shifted], with a
grid of $64\times 256\times 16$ nodes in $\{x,y,s\}$, 
and with node spacings $h_x=h_y=20\pi\rs/64$ and $h_s=2\pi qR/16$.
Nominal parameters corresponding to a typical plasma edge in the
L-mode of tokamak operation are
$$\bhat = 2 \qquad \muhat = 5 \qquad C = 7.65
  \qquad \wcv = 0.05 \qquad \epss = 18350 \qquad \shat=1
	\eqno\eqn
$$
roughly reflecting physical parameters:
$$ n_e = 4.5\times 10^{13}\invcc \qquad
   T_e = 80\eV \qquad B = 2.5\tesla 
	\eqno\eqn
$$$$
   R = 165\cm \qquad \Lpp = 3.65\cm \qquad q = 3
	\eqno\eqn
$$
This is rather strongly collisional (standard parameter $\nu_*=40$), 
but because $C\wcv<1$ it is still
well within the drift wave regime [\bdmode].

\midinsert
% \vskip -10 pt
$$\psboxto(15 true cm;0cm){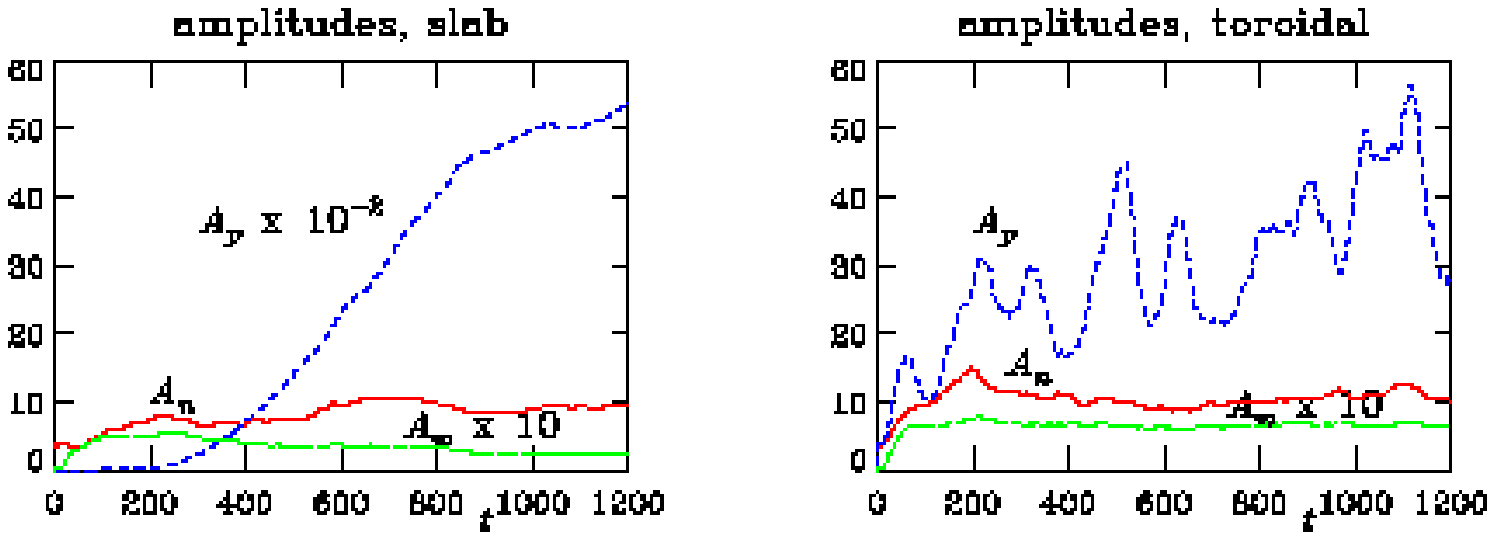}$$
{\hfill\vbox{\hsize=12cm \baselineskip 13 pt \noindent
{\secfnt Figure \figdgdw.} 
Time traces of the squared amplitudes of $\phifl$ (Ap), $\pefl$ (An),
and $\vorfl$ (Aw).  Due to the disparate $\kpp$ factors, Ap tracks
mostly the flows and Aw mostly the turbulence.  The basic slab case
shows initial saturation and then weakening of the turbulence as the
flow amplitude rises.  The basic toroidal case shows persistent
saturation, as the flow amplitude remains low.
}\hfill}

\vskip 20 pt
\endinsert

We refer to the cases with $\wcv=0$ and $0.05$ as the basic slab and
toroidal cases, respectively.  The time traces for these cases
appear in 
Fig.~\figdgdw.  The squared amplitudes are shown for $\phifl$
(Ap), $\pefl$ (An), and the vorticity $\ddpp\phifl$ (Aw).  When strong
zonal flow layers appear, they dominate the Ap signal because of the
lack of $\kpp$ factors.  The Aw signal by contrast, with four additional
$\kpp$ factors, mostly tracks the turbulence.
For the basic slab case, 
the turbulence saturates in the time range
$200<t<400$, after which it is ground down by the rise of the overall
flow levels; Ap grows to large values, and Aw correspondingly decreases.  
For the  basic toroidal case, the saturation
occurs at roughly the same time scale, but the Ap curve saturates
unsteadily at a much lower 
level, smaller by about two orders of magnitude as in the slab case,
reflecting the flows which are simply part of the turbulence.  The
turbulence saturates and maintains its level, close to the basic gyro
Bohm transport.  All time traces reflect this saturated state.
We therefore find that the spin up and suppress scenario operates
moderately well in slab geometry but not at all in toroidal geometry
(for the same basic result in models including both temperatures see
[\gyro]).

The effort to explain this perhaps startling result forces systematic
address of the various toroidal effects, all of which (in this model)
operate through the curvature terms.  There are two basic
effects in the curvature operator
$\kappacv$: the interchange dynamics itself, and the geodesic
curvature. 
Pure interchange dynamics operates on the $k_y\ne 0$ part, through
$\kappacv^y\ppy{}$.  The geodesic curvature effect is in
$\kappacv^x\ppx{}$.  To test directly for the geodesic
curvature one must separate $\kappacv^x$ out for the $k_y=0$ part
and leave the $\kappacv^y$
pure interchange effect alone.  The reason for suspecting the geodesic
curvature is that the zonal flow effects lie in the $k_y=0$ part, for
which the pure interchange effect vanishes due to the vanishing
$\ppy{}$.  

{\sc The Geodesic Curvature Effect.}  The basic mode of oscillation
involving the geodesic curvature is the classic MHD geodesic acoustic
oscillation [\geodesic], which represents simple coupling between the
pressure and vorticity through the geodesic curvature $\kappacv^x$.  The
pressure part of this is a sideband with parallel wavenumber $\kpl qR=1$ 
which presents itself as
an interim free energy source to the turbulence.  The potential part is
the zonal flow; both are axisymmetric ($k_y=0$).  Due to the strong
direct cascade tendency, the nonlinear ExB pressure advection,
$\vedl\pefl$, quickly delivers this free energy back to the turbulence.
Overall, this transfer process acts as a depletion channel for zonal
flow energy, keeping the zonal flow amplitude at levels comparable to
the turblence.  The loss channel is from the zonal flow $\phifl$ to the
sideband 
$\pefl$ (the $\kappacv^x$ terms in Eqs.~\eqvorfl,\eqpefl), and
then through $\vedl\pefl$ back to the eddies of the turbulence.  The
mutual energy transfer is conservative, so the tendency of the system to
reach equipartition results in a finite population of the zonal flow
mode, but not so large as to overwhelm the turbulence.  We note that the
fact that the geodesic curvature couples all $\kpl$ modes of a given
$k_y$ requires us to keep or remove it for all $k_y=0$ modes as a unit;
otherwise, the resulting model would not conserve energy.

\midinsert
% \vskip -10 pt
$$\psboxto(15 true cm;0cm){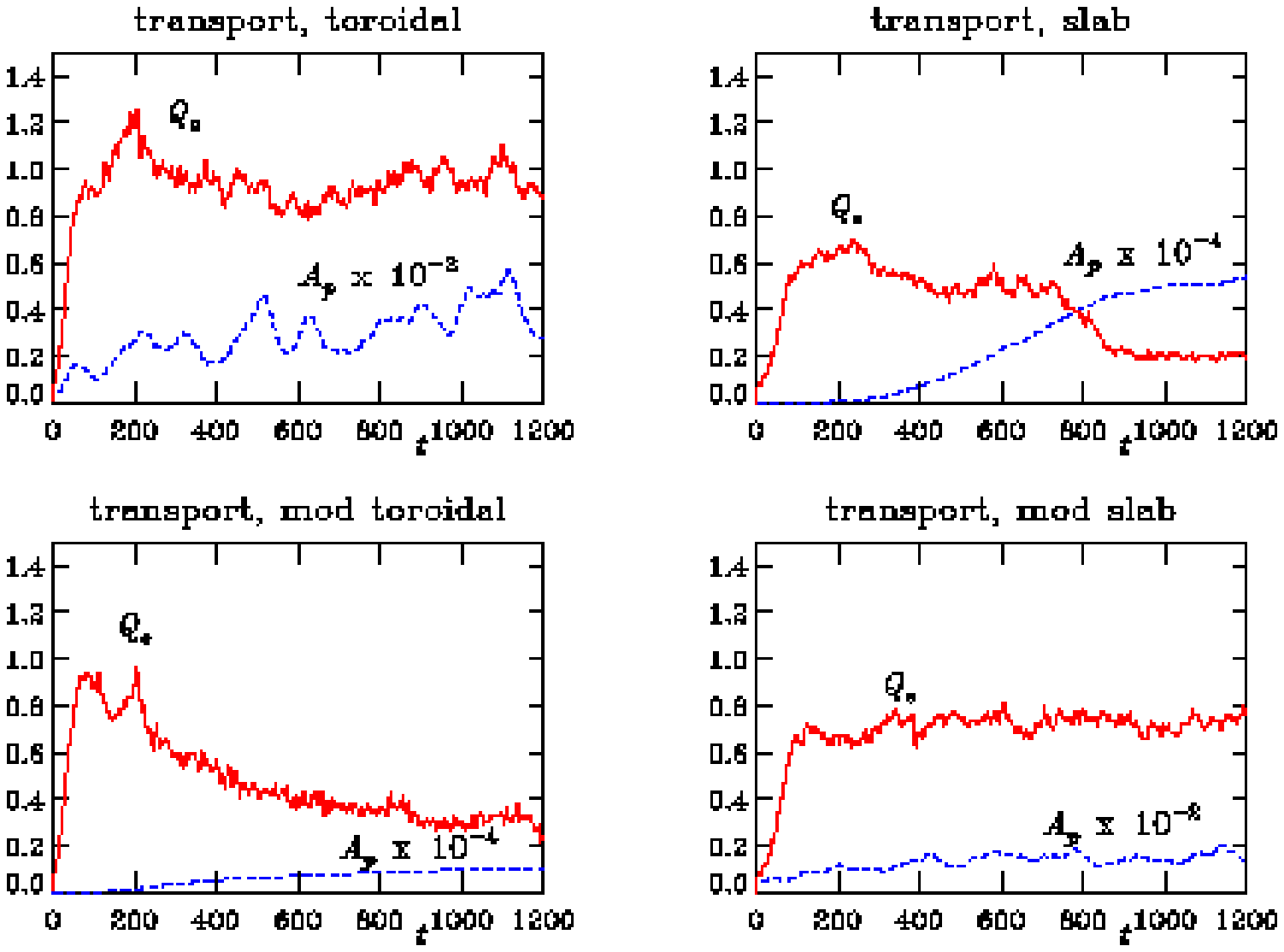}$$
{\hfill\vbox{\hsize=12cm \baselineskip 13 pt \noindent
{\secfnt Figure \figdgdwgd.} 
Time traces of the squared amplitudes of $\phifl$ (Ap) and the transport
(Qe), for the four cases.  The basic toroidal and modified
slab cases reach persistent saturation; both contain the geodesic
curvature effect.  The basic slab and modified toroidal cases
lack this effect and are both ground down by strong, self generated
flow shear.  This test confirms the geodesic curvature effect of
coupling zonal flows to finite $\kpl$ sidebands as the reason the spin
up and suppress scenario does not work in toroidal geometry.
}\hfill}

\vskip 20 pt
\endinsert

A modified toroidal case is constructed by taking $\kappacv^x$
out of the $k_y=0$ part of the basic toroidal case, and a modified slab
case is made by putting $\kappacv^x$ into the $k_y=0$ part of the basic
slab case, thereby isolating the geodesic curvature effects
on the $k_y=0$ part.
The results concerning the turbulence amplitudes and transport are shown
in Fig.~\figdgdwgd.  Two time traces are shown for each case: the
transport (Qe) and the $\phifl$ squared amplitude (Ap).  
We find immediately that the two cases without geodesic curvature
in the $k_y=0$ part are similar, with the flow amplitude rising to
high values, grinding down the transport.  The two cases with the
geodesic curvature
in the $k_y=0$ part are also similar, with the potential amplitude kept at
levels low enough that the turbulence is not suppressed.

\midinsert
% \vskip -10 pt
$$\psboxto(16 true cm;0cm){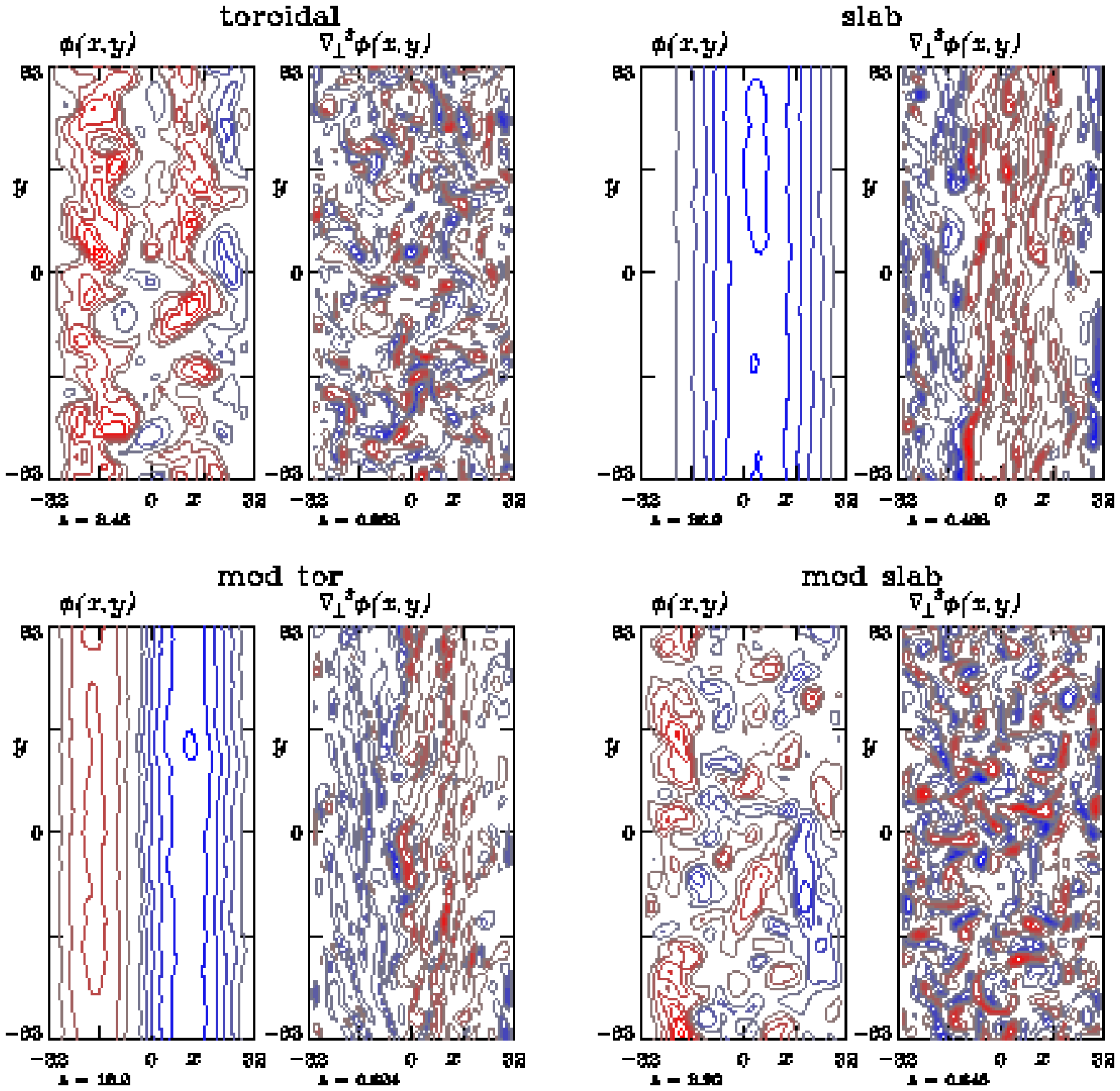}$$
{\hfill\vbox{\hsize=12cm \baselineskip 13 pt \noindent
{\secfnt Figure \figplctr.} 
Morphology of the flows and disturbances ($1/2$ of the $y$-domain is
shown).  The basic slab case shows the strong shear layers in $\phifl$,
and sheets of vorticity ($\vorfl$) stretched in the $y$-direction.  The
basic toroidal case shows visible shear layers in $\phifl$, but they are
of similar magnitude as the turbulence and do not strongly affect the
form of $\vorfl$.  These are the time dependent zonal flow layers
visible in the unsteady $\phifl$ amplitude in Fig.~\figdgdwgd.  The
modified toroidal case appears slablike, while the modified slab case
looks like the basic toroidal case.
}\hfill}

\vskip 20 pt
\endinsert

The morphology of $\phifl$ and $\pefl$ is shown for the four cases in
Fig.~\figplctr.  The basic slab case shows dominance of 
the shear layer in the potential, with the vorticity disturbances
stretched into thin sheets sharply tilted into the $y$-direction.
The basic toroidal case also shows shear layers,  but their vorticity
represents a frequency not larger than that of the basic 
turbulence, which is why a strong amount of suppression does not occur.
The modified toroidal case shows the strong
shear layer of the basic slab case, and the modified slab case
shows
the structure of the basic toroidal case.  The shear levels of
these weaker flows are comparable to the dynamical frequencies of the
turbulence (about $0.1 c_s/\Lpp$).
The Ap curves
for these two cases with weaker flows show those flows to be
short-lived, comparable to 
the correlation time of the turbulence (about $6\Lpp/c_s$).  These are
the zonal flows which remain as part of the turbulence, leading in fact
to moderate suppression
but allowing it to
remain at a robust amplitude.

\midinsert
% \vskip -10 pt
$$\psboxto(13.1 true cm;0cm){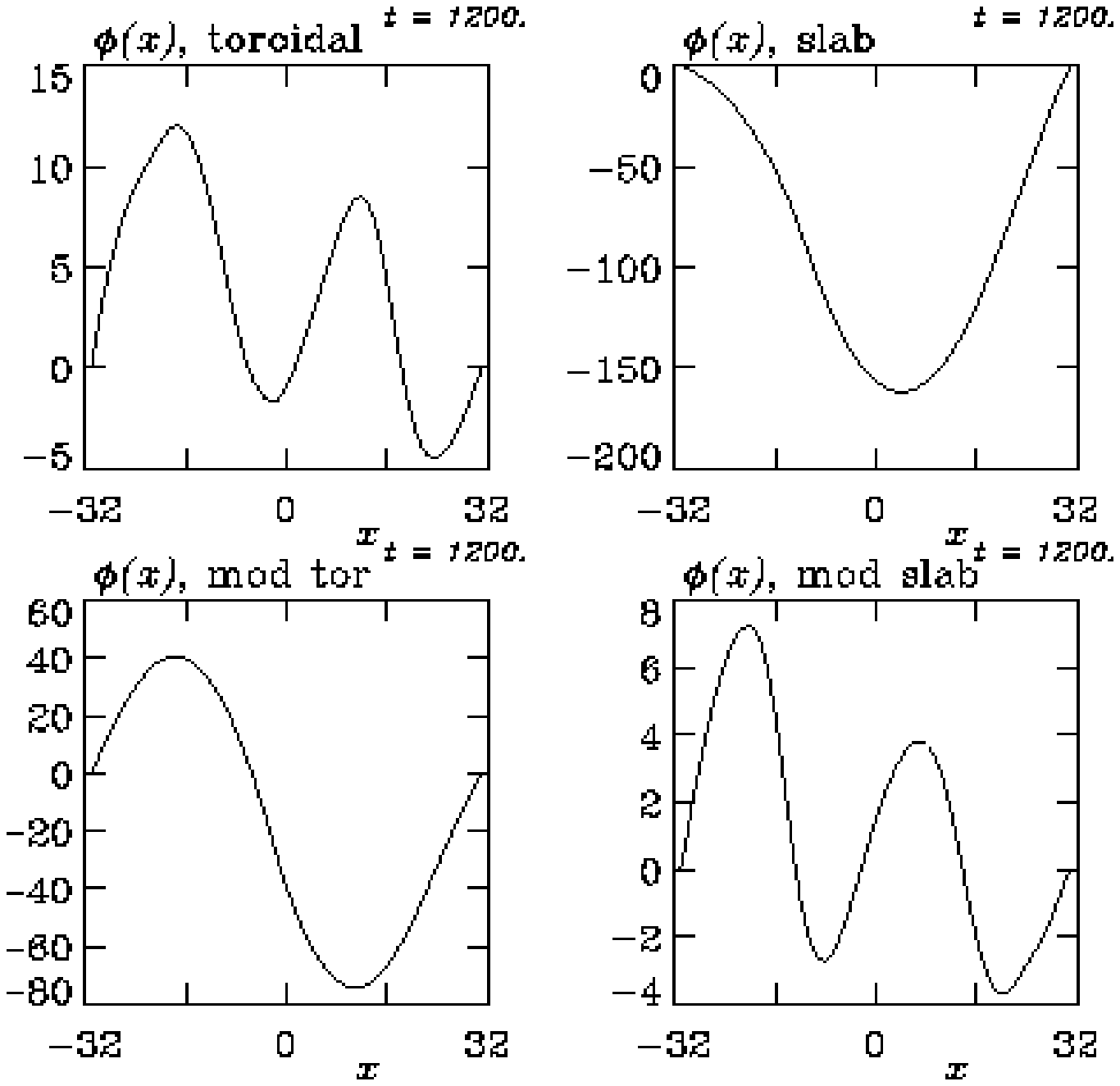}$$
{\hfill\vbox{\hsize=12cm \baselineskip 13 pt \noindent
{\secfnt Figure \figplprofgd.} 
Snapshots of the zonal flow profiles for the four cases.
The basic toroidal and modified slab cases show the weak, time dependent
zonal flows which are part of the turbulence.  The basic slab and
modified toroidal cases show the strong, self generated shear layers
which suppress the turbulence.
}\hfill}

\vskip 20 pt
\endinsert

The instantaneous profiles (the zonal flow mode, $k_y=\kpl=0$) of
$\phifl$ are shown for the four cases in Fig.~\figplprofgd, in the same
arrangement as for the time traces.  The flow shear of these is strong
or weak according to whether the geodesic curvature is absent 
or present in the $k_y=0$ part, respectively.
A rough guide of whether these sheared flow layers are able to suppress
the turbulence is given by what can be called the ``diamagnetic flow
shear level'' given by 
$$\Omega_D=v_D/\Lpp
	\eqno\eqn$$
where $v_D=cT_e/eB\Lpp$ is the
diamagnetic velocity.  The level of shear in the $\phifl$ profiles
(actually given by the vorticity profile) is well below this for the two
unsuppressed cases (basic toroidal and modified slab), and well above
this for the other two cases.
(ALT: show vor profiles, note diag shear is $\rho_*$ in n.u., while turb
omega is about 0.1) 

\midinsert
% \vskip -10 pt
$$\psboxto(15 true cm;0cm){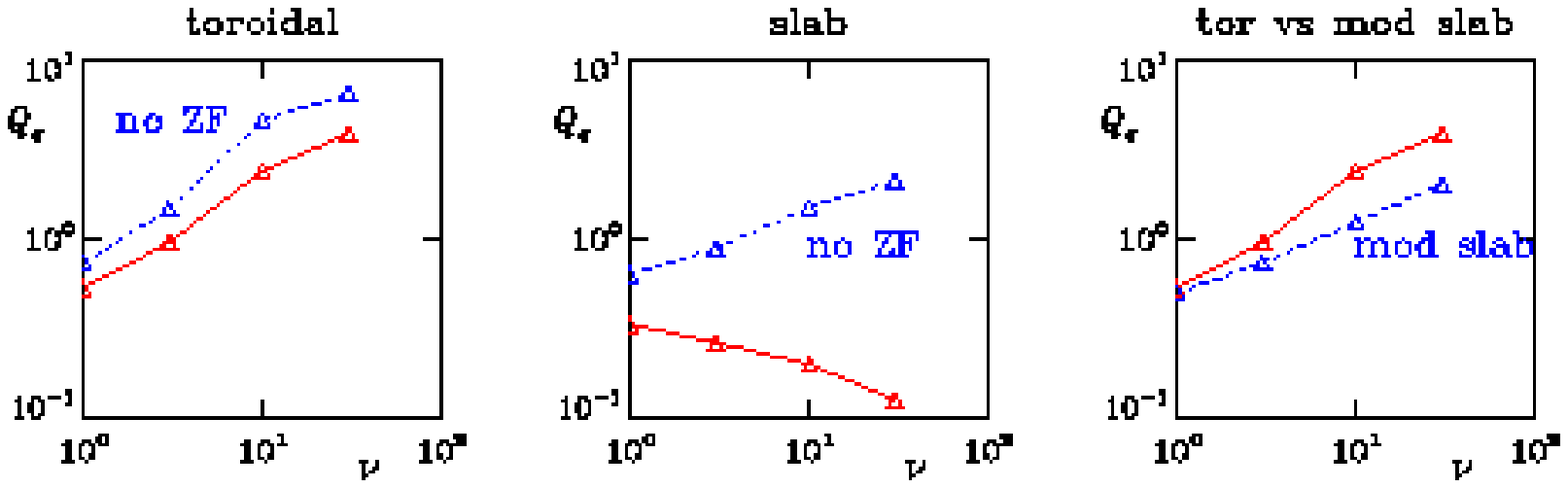}$$
{\hfill\vbox{\hsize=12cm \baselineskip 13 pt \noindent
{\secfnt Figure \figtransport.} 
Transport trend versus collisionality.  The drift wave regime extends to
$\nu=10$, at which $C\wcv\approx 1$, and for these parameters the
standard $\nu_*$ is 136.  The basic toroidal and slab cases are compared
to companion runs in which the zonal flow drive is removed.  The time
dependent zonal flows are the difference in the toroidal case; the self
generated shear layers, in the slab case.  The comparison between the
basic toroidal and modified slab cases shows the role of pure
interchange forcing; the difference at $\nu=3$ ($C=7.65$) is about 20
percent.
}\hfill}

\vskip 20 pt
\endinsert

The transport results for various collisionality are summarised in
Fig.~\figtransport\ (note $C=2.55\nu$).  The basic toroidal and slab
cases show similar trends if zonal flows are eliminated entirely by
removing the flux surface average of $\vedl\vorfl$.  The fluctuating
zonal flows provide a slightly reduced transport in the basic toroidal
case, but the strong shear layers in the basic slab case strongly
suppress the turbulence, even more so for larger $C$.  The modified slab
case is much like the basic toroidal case, showing the effects of
geodesic curvature to inhibit the strong shear layers, leaving the
fluctuating zonal flows and the pure interchange effects intact.  The
small difference between those two cases is the pure interchange effect,
incidentally showing that direct interchange drive in toroidal geometry
is but a small perturbation on an existing drift wave mode structure.

% zzz zfenergy

The coupling of zonal flows to geodesic acoustic oscillations can be
further demonstrated through the energy theorem satisfied by the zonal
flows and the pressure sidebands.  Let $\avg{\cdots}$ denote the
flux surface average and note that it commutes with $\ppx{}$ but
annihilates $\ppy{}$.  Let $\LBR\cdots\RBR$ further denote the average
over $x$.  The zonal flow potential is $\avg{\phi}$, the zonal flow is
$\avg{v^y}=\avg{\ppx{\phi}}$, the zonal vorticity is
$\avg{\Omega}=\avg{\ppx{v^y}}$, and the zonal flow energy is
$\LBR\avg{v^y}^2\RBR$.  Through the 
geodesic curvature the zonal flow is coupled to the Pfirsch-Schl\"uter
Alfv\'en mode and then again to modifications in the background
pressure (assuming unit diagonal metric, and neglecting sound waves,
magnetic nonlinearities, and sidebands with $\abs{\kpl qR}>1$):
$$\ptt{}\LBR\half\avg{v^y}^2\RBR
	= \LBR\avg{\Omega}\avg{v^x v^y}\RBR
		- \wcv\LBR\avg{p_e\sin s}\avg{v^y}\RBR\eqno\eqn$$
$$\eqalign{\ptt{}\LBR\avg{p_e\sin s}^2\RBR
	& = 2\LBR\avg{\pxx{p_e}\sin s}\avg{Q^x\sin s}\RBR
		+ \wcv\LBR\avg{p_e\sin s}\avg{v^y}\RBR
	\cr&\qquad{}
%$$$$\hskip 4 cm{}
		- \wcv\LBR\avg{p_e\sin s}\avg{\pxx{p_e}}\RBR
%		- 2\LBR\avg{p_e\sin s}\avg{(\Jpl-\upl)\cos s}\RBR
		- 2\LBR\avg{p_e\sin s}\avg{\Jpl\cos s}\RBR
	\cr}\eqno\eqn$$
$$\eqalign{\ptt{} & \LBR\bhat^{-1}\avg{B^y\cos s}^2
		+\muhat\avg{\Jpl\cos s}^2\RBR
	\cr&\qquad\qquad{}
	= 2\LBR\avg{\Jpl\cos s}\avg{(p_e-\phi)\sin s}\RBR
		- 2C\LBR\avg{\Jpl\cos s}^2\RBR
	\cr}\eqno\eqn$$
$$\eqalign{\ptt{}\LBR\avg{v^y\sin s}^2\RBR
	& = 2\LBR\avg{\Omega\sin s}\avg{v^x v^y\sin s}\RBR
	\cr&\qquad{}
%$$$$\hskip 4 cm{}
		+ \wcv\LBR\avg{\pxx{p_e}}\avg{\phi\sin s}\RBR
		+ 2\LBR\avg{\phi\sin s}\avg{\Jpl\cos s}\RBR
	\cr}\eqno\eqn$$
$$\ptt{}\LBR\half\avg{p_e}^2\RBR
	= \LBR\avg{\pxx{p_e}}\avg{Q^x}\RBR
		- \wcv\LBR\avg{\pxx{p_e}}\avg{(\phi-p_e)\sin s}\RBR\eqno\eqn$$
where $Q^x = p_e v^x$ is the pointwise transport and
$B^y=-\beta\ppx{\Apl}$ is the field disturbance.  If time averages are
taken, the left sides of these equations vanish and the right sides
become balances between drive, transfer, and depletion mechanisms.  The
drive for the zonal flow is the zonally averaged Reynolds stress
$\avg{v^x v^y}$, correlated with the zonal vorticity.  The depletion
mechanism is geodesic transfer to the Pfirsch-Schl\"uter sideband
$\avg{p_e\sin s}$, and 
the depletion for that is the nonlinear transfer of free energy back to
the turbulence (which requires nonadiabatic electrons, enabled
by the finite $\bhat$).  The flow sideband $\avg{v^y\sin s}$ is
controlled by relaxation of the Pfirsch-Schl\"uter current
$\avg{\Jpl\cos s}$.  The proper geodesic acoustic 
oscillation is that between $\avg{\phi}$ and $\avg{p_e\sin s}$ as
correctly noted in [\geodesic], the only subsystem not strongly affected
by the Alfv\'en dynamics.  We recognise
$\LBR\avg{\ppx{p_e}}\avg{Q^x}\RBR$ as the negative of the main drive of
the self sustained turbulence (cf.\ [\ssdw]), and hence as the
corresponding depletion of the background $\avg{p_e}$.  The profile is
maintained by the damping of the $k_y=0$ part of $\pefl$ to zero at the
boundaries in $x$, which affects both $\avg{p_e}$ and $\avg{p_e\sin s}$.
For the nominal case the zonal
Reynolds stress and geodesic transfer were measured at
$0.858\pm0.359$ and $1.02\pm0.305$, and depletion of the sideband went
through the nonlinearity, the profile maintenance, and the
Pfirsch-Schl\"uter transfer to the background at
$0.470\pm0.431$ and $0.301\pm0.100$ and $0.103\pm0.0741$, respectively,
with all other sideband effects much smaller (all numbers $\times
10^{-2}$).
The nonlinear depletion of $\avg{p_e\sin s}$ functions
because of the pointwise correlation of $Q^x$ with $-\ppx{p_e}$.
Indeed, while either of these two $\sin s$ terms is small in time
average, the average of their product is not, because of this
correlation.  This energetic depletion therefore overwhelms any slight
presence of a nonzero $\avg{Q^x\sin s}$ (Stringer-Taylor effect).
Indeed, the PDF of $\avg{Q^x\sin s}$ was found to be close to Gaussian.
The pressure nonlinearity is therefore a depletion of the sideband and
therefore ultimately of the zonal flow.  The energy flow is
$\pefl\to\avg{\phi}\to\avg{p_e\sin s}\to\pefl$, through the Reynolds
stress, geodesic curvature, and nonlinear flux correlation,
respectively.  The Stringer-Taylor effect is therefore a {\it sink} for
the zonal flow system, not a source as incorrectly reported in
Ref.~[\ksh], whose runs were apparently not taken to complete
statistical saturation and in any case suffer from all the shortcomings
of the drift resistive ballooning paradigm (cf.\ Ref.\ [\bdmode]).

% zzz end zfenergy

{\sc Main Points.}  The principal result of this study is that while the
turbulent Reynolds stress always tends towards a transfer of energy from
small eddies to the larger scale zonal flows (similar $k_x$ but
disparate $k_y$), in toroidal geometry the geodesic curvature couples
the zonal flows to finite-$\kpl$ pressure sidebands, which act as a loss
channel by means of nonlinear advective transfer back to the turbulence.
This prevents large scale, large amplitude zonal flows from
forming and therefore rules out the spin up and suppress scenario for
transport barrier formation, at least due to local (homogeneous) action
of the ExB Reynolds stress.  The ExB shear layers observed in tokamak
edge transport barriers [\wagner] must therefore come from some other
mechanism, most likely having to do with the neoclassical equilibrium.
Two recent proposals are a generalised ion orbit loss mechanism
[\shaing,\heikkinen], and the generation of a large parallel flow and its
equilibrium electric field profile by coupling to the open field line
regions [\rozhansky].  For the core regions the electron response is
more adiabatic and electrostatic, so that current results on core zonal
flows [\zlin] are not affected.

{
\parindent 0 pt
\frenchspacing
\parskip=10pt plus 1pt minus 1pt
\def\ref##1.##2\par{\par\hangindent 15pt [##1]##2}
\par\section{References}
\ref\hasmim.
A. Hasegawa and K. Mima, \prl{39} (1977) 205; \pf{21} (1978) 87.

\ref\wakhas.
M. Wakatani and A. Hasegawa, \pf{27} (1984) 611.

\ref\dalfloc.
B. Scott, \ppcf{39} (1997) 1635.

\ref\focusdw.
B. Scott, \njp{4} (2002) 52.

\ref\zfterry.
Reviewed by P. W. Terry, \physp{7} (2000) 1653.

\ref \biglari.
H. Biglari, P. Diamond, and P. W. Terry, \pfb{3} (1991) 1.

\ref\diamondkim.
P. Diamond and Y. Kim, \pfb{3} (1991) 1626.

\ref\kraichnan.
R. Kraichnan, \pf{10} (1967) 1417.

\ref\sfdw.
B. Scott, \ppcf{34} (1992) 1977.

\ref\hahm.
T. S. Hahm, M. A. Beer, Z. Lin, G. W. Hammett, W. W. Lee, and
W. M. Tang, \physp{6} (1999) 922.

\ref\zlin.
Z. Lin, T. S. Hahm, W. W. Lee, W. M. Tang, R. B. White, 
{\it Science} 281 (1998) 1835; also R. Sydora (get ref from Dimits).

\ref\wagner.
H-Mode discovery: F. Wagner \etal, \prl{49} (1982) 1408;
H-Mode morphology: P. Gohil, K. H. Burrell, E. J. Doyle, R. J. Groebner,
J. Kim, and R. P. Seraydarian, \nf{34} (1994) 1057.

\ref\diamond.
P. H. Diamond, Y.-M. Liang, B. A. Carreras, and P. W. Terry,
\prl{72} (1994) 2565.

\ref\varenna.
B Scott, ``Physics of Zonal Flows in Drift Wave Turbulence,'' in
Theory of Fusion Plasmas (Editrice Compositori, Bologna, 2000,
J. Connor, O. Sauter, and E. Sindoni, eds), p. 413; cf. also
B. Scott, ``Recent Results Concerning Local and
Global Computation of Self-Consistent Transport Scenarios,'' 
\iaea{1994}{15}, Seville (IAEA, Vienna, 1996), Vol. 3, p. 447.

\ref\shifted.
B. Scott, \physp{8} (2001) 447.

\ref\fluxtube.
B. Scott, \physp{5} (1998) 2334.

\ref\brag.
S. I. Braginskii, \revpp{1} (1965) 205.

\ref\bdmode.
B. Scott, preprint {\tt arXiv:physics.plasm-ph/0207126}, submitted to \physp{}.

\ref\gyro.
B. Scott, \physp{7} (2000) 1845.  

\ref\geodesic.
N. Winsor, J. Johnson, and J. Dawson, \pf{11} (1968) 2448.

\ref\ssdw.
B. Scott, \prl{65} (1990) 3289; \pfb{4} (1992) 2468.

\ref\ksh.
K. Hallatschek and D. Biskamp, \prl{86} (2001) 1223.

\ref\shaing.
K. C. Shaing and E. C. Crume, Jr., \prl{63} (1989) 2369.

\ref\heikkinen.
J. A. Heikkinen, T. P. Kiviniemi, and A. G. Peeters, 
\prl{84} (2000) 487.

\ref\rozhansky.
D. Morozov, V. Rozhansky, J. Herrera, and T. Soboleva,
\physp{7} (2000) 1184.

\par

}

\end